\documentclass[intlimits,twoside,a4paper]{article}

\usepackage[cp1251]{inputenc}

\usepackage{mathrsfs}
\usepackage[eqsecnum]{cmpj3}
\issue{2023}{26}{4}{43704}
\doinumber{10.5488/CMP.26.43704}
\title[Optical and thermal effects in the neighborhood of the spherical layered nanoparticle]%
{Optical and thermal effects in the neighborhood of the spherical layered nanoparticle of the ``metallic core -- J-aggregate shell'' structure%
}
\author[A. V. Korotun, N. A. Smirnova, V. I. Reva, I. M. Titov, G. M. Shilo]{A. V. Korotun\orcid{0000-0003-4165-2788}\refaddr{label1,label2}\thanks{Corresponding author: \email{andko@zp.edu.ua}.},
        N. A. Smirnova\orcid{0000-0002-6514-8562}\refaddr{label1},
        V. I. Reva\orcid{0000-0002-3265-1735}\refaddr{label1},
        I. M. Titov\refaddr{label3},
        G. M. Shilo\orcid{0000-0002-5020-6707}\refaddr{label4}}
\addresses{
\addr{label1} National University Zaporizhzhia Polytechnic, 64 Zhukovsky Str., Zaporizhzhia, 69063, Ukraine
\addr{label2} G. V. Kurdyumov Institute for Metal Physics of National Academy of Sciences of Ukraine, 36 Academician Vernadsky Blvd., Kyiv, 03142, Ukraine
\addr{label3} UAD Systems, 84 Olexandrivska Str., 69002, Zaporizhzhia, Ukraine
\addr{label4} Zaporizhzhia National University, 66 Zhukovsky Str., 69600, Zaporizhzhia, Ukraine
}


\date{Received July 02, 2023, in final form October 01, 2023}

\begin{document}

\maketitle

\begin{abstract}
The relations for the polarizability of the metallic nanoparticles, coated with the shell of cyanine dyes, are obtained in the article. The frequency dependencies for light absorption and scattering efficiencies, the heating of the composite nanoparticle and the electric field amplification in its neighborhood are studied. It is established that all the dependencies have three maxima which correspond to the frequencies of hybrid plasmon-exciton resonance. It is shown that an increase in content of metal in the nanoparticle causes a blue shift of the maxima from the visible part of the spectrum and a red shift of the maximum from ultraviolet frequency range. The issue of application of metal-organic nanoparticles in nanomedicine, in particular for the photothermal therapy of malignant neoplasms is studied.
%
%
%\keywords Up to six keywords (\href{https://physh.aps.org/browse}{Physics Subject Headings})
\printkeywords
%
%\pacs 78.67.Sc, 79.60.Jv, 82.33.Pt, 82.35.Np, 82.65.+r
\end{abstract}

\section{Introduction}

In the recent decade the surface plasmons in metallic nanostructures and the interaction between the surface plasmons and quantum emitters, for example, quantum dots, as well as cyanine dye molecules are under active research~\cite{B1,B2,B3,B4,B5}. Such nanostructures make it possible to control the interaction between light and substance at the subwavelength scale. Their possible applications are associated with optical sensing, photon information processing~\cite{B6,B7}, and detecting ultra-low concentration of substances~\cite{B7++}.

It should be pointed out that the assembly of dyes on plasmonic nanostructures can result in the formation of hybrid systems~\cite{B8,B9}. At the same time, the most suitable organic components are cyanine dyes characterized by the presence of aggregated forms of molecules, such as J-aggregates~\cite{B10}, which demonstrate a narrow absorption peak shifted into the long wavelength region with respect to the absorption spectrum of the monomeric form~\cite{B11,B12}. Moreover, J-aggregates have high optical absorption coefficient (larger than $10^{6}$~cm$^{-1}$)~\cite{B13}. The mentioned absorption features of J-aggregates are fundamentally important for their use in photodetectors, which require characteristic wavelength of incident radiation~\cite{B11}. The unusual optical properties of J-aggregates are also known due to exciton transfer and, hence, they can be used in such applications as organic field-effect transistors and solar cells~\cite{B14}.

An interaction between plasmons in metallic nanostructures and excitons of organic dyes attract attention due to unique properties of hybrid structures, obtained by using two components. For example, hybrid structures are characterized, on the one hand, by high tunability of spectral and spatial properties, similar to plasmonic nanoparticles; on the other hand, by high optical nonlinearity and light-emitting properties, similar to dye excitons~\cite{B15}. The study of plasmon-exciton interaction makes it possible not only to improve the understanding of the nature of the interaction between light and substance, but also becomes the basis for the development of new sensing methods, devices that use solar energy and for the development of new methods of malignant tumors therapy~\cite{B16,B17,B18,B19}.

The related physical phenomenon such as heat dissipation of nanoparticles under the excitation of hybrid resonance is also of great interest~\cite{B36}. The heat dissipation can become the key factor for multiple thermonanoplasmonics applications, in particular, for the rapidly developing fields of nanomedicine~\cite{B37}, nanobiology~\cite{B38} and nanochemistry~\cite{B39,B40} although the heat dissipation at the nanoscale is harmful in some cases. For example, the strong optical absorption and the subsequent radiation-free energy scattering give an opportunity to apply metallic and composite nanoparticles in photothermal therapy of malignant tumors. The possibility of using the spherical metallic nanoparticles, covered with graphene shell, for these purposes was studied in the work~\cite{B41}. The heating in the neighborhood of spherical bimetallic nanoparticle was estimated in works~\cite{B42,B43}.

Plasmonic photothermal therapy is a potentially favorable alternative to the conventional therapy methods for localized tumors, such as chemotherapy, radiation therapy and surgery.
Currently, the photothermal properties of metallic and composite nanoparticles are widely used not only to kill cancer cells~\cite{B37}, but also for the selective destruction of antibiotic-resistant bacteria~\cite{B44}, HIV treatment~\cite{B45}, in coronavirus outbreak control --- for real-time and tag-free detection of viral sequences~\cite{B46}. The other applications are aimed at the creation of the antimicrobial and antiviral compositions based on nanoparticles, which are not only suitable for disinfecting air and surfaces, but are also effective for reinforcing personal protective equipment, such as face respirators~\cite{B47}.

Moreover, the application of thermoplasmonics is possible in such areas as energy~\cite{B48}, solar and thermal energy collection~\cite{B49}, nanofluidics~\cite{B50}, nanofabrication~\cite{B51}, heat-effected magnetic recording (HAMR) for data storage~\cite{B52}.

It is known that plasmon-exciton interaction can appear in different modes: weak and strong.
When molecule is situated near plasmonic structures, one can observe an increase in absorption and emission rates of dye molecules which is due to extreme concentration of light energy near the surface of metallic nanostructures due to the amplification of the local electric field. This phenomenon consists in a weak interaction~\cite{B20}.

In the case of strong plasmon-exciton interaction, a close location of plasmon and exciton is cha\-rac\-terized by equal excitation levels, which leads to a new hybrid state, known as plexciton, and to the splitting of the excited levels~\cite{B21}. Such interaction manifests itself in the form of formation of a spectral gap in the exciton absorption band and in the form of two peaks on either side of the gap in the absorption and fluorescence spectra~\cite{B20}.

The formation of plexcitons was studied on the plane metallic surfaces with the propagating surface plasmon-polaritons~\cite{B22,B23,B24}, as well as under the excitation of localized surface plasmonic resonances~(SPR) of metallic nanoparticles in the solution~\cite{B25}. An interest to plexcitons is caused by the possibility of controlling the properties of substance by influencing the relation ``light--substance'' and by increasing the efficiency of energy and electron transfer~\cite{B26,B27}, as well as by reducing the interaction with environment~\cite{B28}. The listed possibilities are important for applications dealing with the localization of artificial light, sensors, and also in photonics~\cite{B29,B30,B31}. Plexcitonic materials, in which the plasmonic component is colloidal nanoparticle, are of great interest since they are easily tuned, scaled, and they can be easily synthesized with the help of cheap ``wet'' chemistry techniques~\cite{B32,B33}. The majority of the cases of observation of plexcitons in nanoparticles, mentioned in literature, are related to the use of dye aggregates as the molecular exciton component, since the interaction with plasmon increases by a larger dipole moment of transition, which is facilitated by aggregation~\cite{B34,B35}.

It should be noted that in~\cite{B35+} the optical properties of ultrathin metal nanofilms coated with a layer of J-aggregate molecules were studied, and in~\cite{B35++} the effects of plasmon-exciton interaction during the absorption and scattering of light by spherical nanoparticles consisting of a metal core as well as the J-aggregate shells were considered. However, studies of the efficiencies of light absorption and scattering, heating and field enhancement in the vicinity of a hybrid nanoparticle have not been carried out.

Thus, the study of optical and thermal phenomena in the neighborhood of composite metal-organic nanoparticles is an urgent task from both theoretical and practical points of view.

\section{Theoretical concepts and mathematical model}

\subsection{Plasmonic heating in the neighborhood of the nanoparticle}

\begin{figure}[htb]
\centerline{\includegraphics[scale=1]{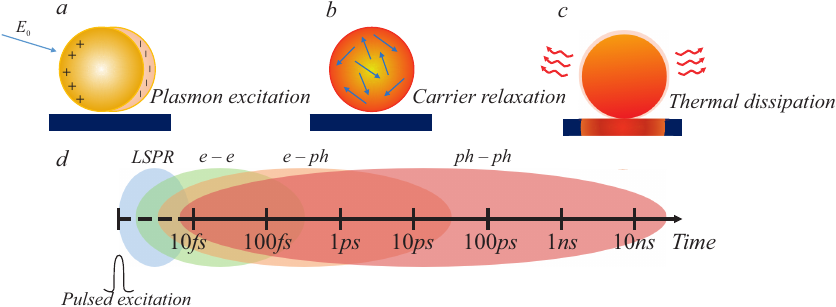}}
\caption{(Colour online) Schematic illustration of photoexcitation of the localized SPR (a); relaxation of photoexcitation of the localized SPR (b); release of thermal energy into the environment (c); energy exchange involved in the photothermal response of the nanoparticle on a logarithmic time scale (d).} \label{fig1}
\end{figure}

Localized surface plasmonic resonance is an effective technique of energy retention in the neighborhood of nanoparticles, leading to a sharp increase both in the absorption and scattering cross-sections~\cite{B53}. While the scattering cross-section characterizes the amount of energy reradiated by the nanoparticle in the form of light, the absorption cross-section determines that part of light energy which is reradiated as heat~\cite{B54}. The advantage of converting light into heat provides access to an innovative way of studying and controlling thermal phenomena at the nanoscale. In order to better understand these processes, it is necessary to take into account that the optical properties of the studied nanoobjects are determined by a number of energy exchanges, each characterized by the characteristic time scale (figure~\ref{fig1})~\cite{B53}.

When interacting with an incident light wave, nanoparticles acquire energy by absorbing photons. If the frequency of the incident radiation is close to the frequency of the localized SPR, there is a resonant connection of the electromagnetic wave with the resulting oscillations of the electron cloud (figure~\ref{fig1}a); this connection appears as collective, coherent dipole oscillations of conduction zone electrons. The result is the global nonequilibrium state. To restore the internal thermal equilibrium of these electrons, there is a redistribution of energy due to electron-electron collisions within the quasi-free electron gas. This process takes place on a time scale of approximately $10~\rm{fs} \leqslant \tau \leqslant 100~\rm{fs}$. Subsequently, the energy of the hot carriers is redistributed during the relaxation process (figure~\ref{fig1}b) due to the scattering associated with the electron-phonon interaction in the time interval $100~\rm{fs} \leqslant \tau \leqslant 1~\rm{ps}$. In the last step, thermal energy is transferred to the interface via phonon-phonon collisions on a time scale of $10~\rm{ps} \leqslant \tau \leqslant 10~\rm{ns}$. This process (thermal dissipation, figure~\ref{fig1}c) causes the cooling of the nanoparticle, which releases heat into the environment, causing the rise of its temperature. The dynamics of this process strongly depend on the thermal properties of the environment~\cite{B55}.

The effect of such energy transfer sequence (figure~\ref{fig1}d) is that the internal energy of the electron gas, after the absorption of photon energy, passes through the following stages: 1) rapid and resonant increase due to localized SPR; 2) electron-electron scattering (athermal mode); 3) electron-phonon scattering; 4)~phonon-phonon scattering (heat transfer into the environment).

\subsection{Cyanine dyes, their molecular forms and optical properties}

\begin{figure}[htb]
\centerline{\includegraphics[scale=0.7]{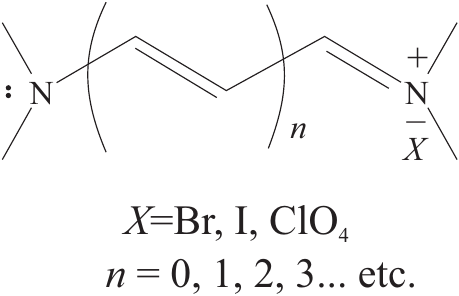}}
\caption{Molecular structure of the cyanine dye~\cite{B57}.} \label{fig2}
\end{figure}

Cyanine dye molecules are two nitrogen heterocyclic rings connected by the polymethine chain, as shown in figure~\ref{fig2}. The polymethine chain contains carbon atoms in sp$^2$-hybridization, which form a strongly bonded chain with the positive charge on nitrogen due to the displacement of $\pi$-electrons~\cite{B56}. Cyanine dyes are classified according to the number of methyl groups $(n)$ in the chain between the two ring systems and the nature of the present ring part.

The spectral properties of cyanine dyes depend largely on the number of methyl groups in the polymethine chain~\cite{B53}. Each lengthening of the chromophore by one vinylene part (-CH=CH-) causes the bathochromic shift of about 100--130~nm~\cite{B54,B55}.

Dye molecule usually carries a positive charge distributed along the polymethine chain. Due to the uniform distribution of $\pi$-electrons in the polymethine chain, these dyes have narrow absorption bands with large extinction coefficients (of the order of $10^5\cdot l \cdot \rm{mol}^{-1}\cdot\rm{cm}^{-1}$) in the spectral range from 340 to 1400 nm~\cite{B58,B59}.

As a rule, the absorption spectra of cyanine dyes contain only one short-wavelength oscillation maximum as a characteristic shoulder at the edge of the main band. Cyanine dyes differ from other classes of organic dyes by their capability to actively absorb light and form molecular aggregated forms, which is a striking example of self-organization in organic substances~\cite{B60}.

\begin{figure}[htb]
\centerline{\includegraphics[scale=0.8]{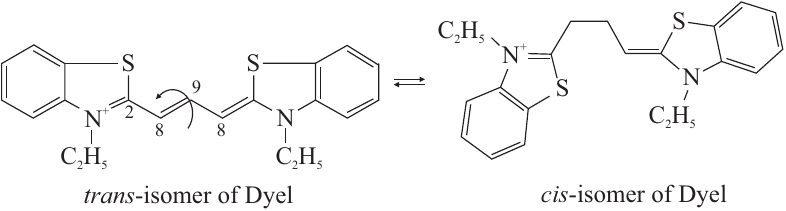}}
\caption{Molecule structures of all-trans- and cis-isomers of the dye 3-ethyl-2-\{[1E(Z),3Z]-3-[3-ethylbenzothiazol-2(3H)-yliden]prop-1-enyl\}benzothiazol-3-ium~\cite{B59}.} \label{fig3}
\end{figure}

In weakly concentrated solutions and thin layers of such solutions, the basic molecular form is the all-trans isomer. Since the cyanine dye molecule is not rigid, various oscillations and rotations of individual fragments are possible in it, including torsion around various bonds of the polymethine chain. Also in cyanine dye molecules, there is a possibility of isomerization due to rotation around the C=C bond of the polymethine chain of the molecule~\cite{B58}. One of the main ways of relaxation of the state energy $S_1$ of cyanine dyes in low-viscosity solvents is usually the transition of all-trans-form into cis-isomers, which takes place by rotating the fragment of the dye molecule around one of the bonds of the polymethine chain by $\sim 180^{\circ}$ (figure~\ref{fig3}). The absorption band of cis-isomers is shifted to the shorter wavelength region with respect to the absorption band of all-trans-isomers. All-trans-isomers of cyanine dyes have high values of the fluorescence quantum yield, while cis-isomers practically do not fluoresce~\cite{B59,B60,B61}.

\begin{figure}[htb]
\centerline{\includegraphics[scale=0.8]{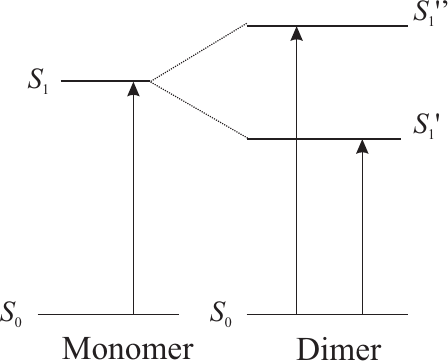}}
\caption{Scheme of energy levels of dye monomers and dimers~\cite{B65}.} \label{fig4}
\end{figure}

An increase of the concentration of dye in solutions leads to the formation of aggregated molecular forms~\cite{B62}. Such aggregates demonstrate noticeable changes in the absorption band compared to the absorption band of monomeric molecules. The simplest cases of the aggregated form of cyanine dye molecule are dimers. Let us consider the spectral manifestation of molecular interaction in dimers. Figure~\ref{fig4} shows a diagram of the energy levels of the monomer and dimer, which shows a splitting of the singlet excited state of the monomer $S_{1}$, resulting in the formation of levels $S_{1}^{\prime}$ and $S_{1}^{\prime\prime}$ of the dimer. As a result of the transitions $S_{0} \rightarrow S_{1}^{\prime}$ and $S_{0} \rightarrow S_{1}^{\prime\prime}$, two bands are observed in the absorption spectrum of the dimer~\cite{B63}. The probability of the electronic transition from the normal state to the excited state, which is split due to the intermolecular interaction, is affected by the value of the angle between the dipoles and the aggregate axis. When the angle is equal to $90^{\circ}$, the probability of the spectral transition to the upper excited state $S_{1}^{\prime\prime}$ is the highest. This transition corresponds to dimers with the parallel arrangement of molecules with respect to each other, whose absorption spectrum is located in the shorter wavelength region with respect to the monomeric form. When the angle between the dipoles and the aggregate axis is equal to $0^{\circ}$, the transition to the lower excited state $S_{1}^{\prime}$ is observed, which leads to the shift of the dimer absorption band to the long-wave region compared to the monomer band~\cite{B64}.

\begin{figure}[htb]
	\centerline{\includegraphics[scale=0.80]{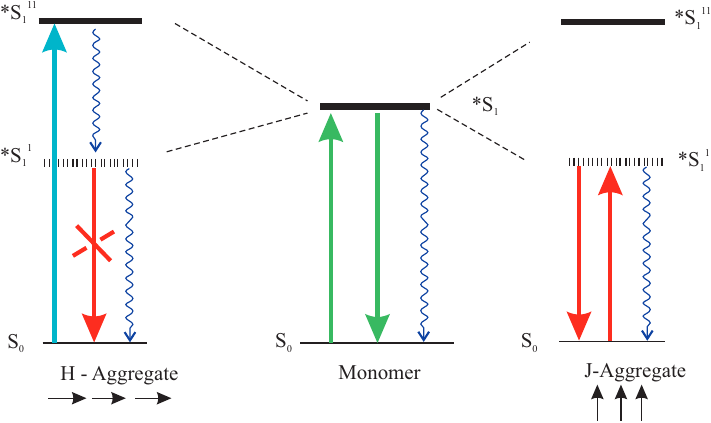}}
	\caption{(Colour online) Excitonic splitting of levels under the formation of dye aggregates~\cite{B67}.} \label{fig5}
\end{figure}

At concentrations of the dye solution above those values at which dimers are formed, two more forms appear: J-aggregates and H-aggregates. Such aggregation of forms is best explained by the exciton theory put forward by McRae and Kash~\cite{B65,B66}. According to the exciton theory (figure~\ref{fig5}), the dye molecule is considered to be a point dipole. The exciton band is formed due to the relation between dipoles in the excited state. The dye aggregation splits its singlet state into two energy levels. When the transition to the lower energy level is allowed, the narrow absorption with the red shift is observed, which is characteristic of J-aggregates. When the transition to the higher energy level takes place, the hypsochrome-shifted absorption band is observed, which is characteristic of H-aggregates~\cite{B56}.

In J-aggregates, the dye molecules form a sliding parallel arrangement (head-to-tail), while in the case of H-aggregates there is a non-sliding parallel arrangement (head-to-head). J-aggregates demonstrate a narrow absorption band and usually result in enhanced fluorescence with very little Stokes shift. H-aggregates, by contrast, ideally are not capable of fluorescing~\cite{B67}.

\subsection{Basic relations}

\begin{figure}[htb]
\centerline{\includegraphics[scale=0.5]{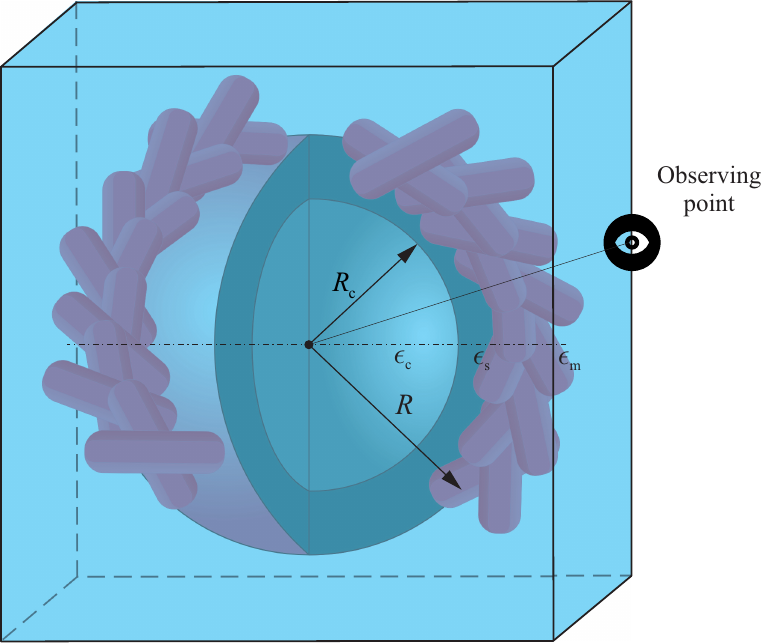}}
\caption{(Colour online) Geometry of the problem.} \label{fig6}
\end{figure}

Let us consider a spherical metallic nanoparticle with the radius $R_c$, covered with the layer of J-aggregate with the thickness $t$, located in the medium with dielectric permittivity $\epsilon_{\rm(m)}$ and which has the total radius $R = R_c + t$ (figure~\ref{fig6}).
The increase in temperature due to the absorption of light energy by the nanoparticle can be estimated as follows~\cite{B69,B70}
\begin{equation}\label{eq1}
  \Delta T = \frac{{{I_0}R}}{{4\lambda }}Q_@^{{\rm{abs}}},
\end{equation}
where $I_0$ is the incident light intensity; $\lambda$ is the environmental thermal conductivity; $Q_@^{\rm{abs}}$ is the efficiency of absorption, which is determined by the relation~\cite{B42}
\begin{equation}\label{eq2}
  Q_@^{{\rm{abs}}} = \frac{\omega }{{\piup {R^2}c}}\sqrt {{\epsilon_{\rm{m}}}} \Im {\alpha _@}.
\end{equation}
In formula (\ref{eq2}) $\omega$ is the frequency of incident electromagnetic wave; $c$ is the light velocity, and $\alpha _@$ is the dipole polarizability of the composite nanoparticle
\begin{equation}\label{eq3}
  {\alpha _@} = {R^3}\frac{{\left( {{\epsilon_{\rm{J}}} - {\epsilon_{\rm{m}}}} \right)\left( {2{\epsilon_{\rm{J}}} + {\epsilon_{\rm{c}}}} \right) + {\beta _{\rm{c}}}\left( {2{\epsilon_{\rm{J}}} + {\epsilon_{\rm{m}}}} \right)\left( {{\epsilon_{\rm{c}}} - {\epsilon_{\rm{J}}}} \right)}}{{\left( {{\epsilon_{\rm{J}}} + 2{\epsilon_{\rm{m}}}} \right)\left( {2{\epsilon_{\rm{J}}} + {\epsilon_{\rm{c}}}} \right) + {\beta _{\rm{c}}}\left( {{\epsilon_{\rm{J}}} - {\epsilon_{\rm{m}}}} \right)\left( {{\epsilon_{\rm{c}}} - {\epsilon_{\rm{J}}}} \right)}},
\end{equation}
where ${\beta _{\rm{c}}} = {\left( {{{{R_{\rm{c}}}} \mathord{\left/
 {\vphantom {{{R_{\rm{c}}}} R}} \right.
 \kern-\nulldelimiterspace} R}} \right)^3}$ is the bulk content of metallic fraction in the composite nanoparticle. The dielectric function of the core and shell materials have the form:
\begin{equation}\label{eq4}
\epsilon_{\rm{c}}\left( \omega  \right) =\epsilon^{\infty } - \frac{\omega_p^{2}}{\omega \left( \omega  + {\mathop{\rm i}\nolimits} {\gamma _{\rm{eff}}} \right)},
\end{equation}
\begin{equation}\label{eq5}
\epsilon_{\rm{J}}\left( \omega  \right) = \epsilon_{\rm{J}}^\infty  + \frac{{f\omega _0^2}}{{\omega _0^2 - {\omega ^2} - {\mathop{\rm i}\nolimits} \omega {\gamma _{\rm{J}}}}}.
\end{equation}
In formulae (\ref{eq4}) and (\ref{eq5}), $\omega_p$ is the plasma frequency; $\epsilon^{\infty}$ is the contribution of the crystal lattice into the dielectric function of the metallic core; $f$ is the reduced oscillator force; $\epsilon^{\infty}_{J}$ is the value of the dielectric permittivity of J-aggregate away from the center of the absorption band; $\omega_0$ is the frequency which corresponds to the center of the band, $\gamma_{\rm{J}}$ is the width of Lorentz contour of J-band. The effective relaxation rate in the nanoparticles is determined by the sum of contributions of the bulk and surface relaxations and the radiation damping~\cite{B41}
\begin{equation}\label{eq6}
\gamma _{\rm{eff}} =\gamma _{\rm{bulk}} + \gamma _{\rm{s}} + \gamma _{\rm{rad}}.
\end{equation}
The bulk relaxation rate is a constant value, the radiation damping for particles of the considered sizes can be neglected, and the surface relaxation rate depends both on frequency and on the size of nanoparticles
\begin{equation}\label{eq7}
\gamma _{\rm{s}} = \mathscr{A}\left( {\omega ,R} \right)\frac{{v_{\rm{F}}}}{{R_{\rm{c}}}},
\end{equation}
where $v_{\rm{F}}$ is the Fermi electron velocity. An effective parameter of coherence loss for the spherical nanoparticles is determined by the expression:
\begin{equation}\label{eq8}
\mathscr{A}\left( \omega, R \right) = \frac{1}{4}{\left( {\frac{{{\omega _{\rm{p}}}}}{\omega }} \right)^2}\left[ {1 - \frac{{2{\nu _s}}}{\omega }\sin \frac{\omega }{{{\nu _s}}} + \frac{{2\nu _s^2}}{{{\omega ^2}}}\left( {1 - \cos \frac{\omega }{{{\nu _s}}}} \right)} \right],
\end{equation}
and ${\nu _s} = {v_{\rm{F}}} \mathord{\left/
 {\vphantom {{{v_{\rm{F}}}} {2R_{\rm{c}}}}} \right.
 \kern-\nulldelimiterspace} {2R_{\rm{c}}}$ is the frequency of individual oscillations of electrons.
The factor of the field amplification near the particle under $r > R$ is determined by the relation
\begin{equation}\label{eq9}
\mathscr{G}\left( r \right) = {\left| 1 + 2\frac{\alpha _{@}}{r^{3}} \right|^2},
\end{equation}
where $r$ is the distance between the center of the composite nanoparticle and the point in which the factor of the field amplification is calculated.

Let us determine approximately the frequencies of hybrid plasmon-exciton resonances in non-dissipative approximation (under the condition $\gamma _{\rm{eff}} = \gamma _{\rm{J}} = 0$) using the condition of the equality to zero of the denominator in the expression (\ref{eq3}):
\begin{equation}\label{eq10}
\left( {\epsilon_{\rm{J}}} + 2{\epsilon_{\rm{m}}} \right)\left( {2{\epsilon_{\rm{J}}} + {\epsilon_{\rm{c}}}} \right) + {\beta _{\rm{c}}}\left( {{\epsilon_{\rm{J}}} - {\epsilon_{\rm{m}}}} \right)\left( {{\epsilon_{\rm{c}}} - {\epsilon_{\rm{J}}}} \right) = 0.
\end{equation}
The expressions (\ref{eq4}) and (\ref{eq5}) have the following form in the mentioned approximation
\begin{equation}\label{eq11}
{\epsilon_{\rm{c}}}\left( \omega  \right) =\epsilon^{\infty } - \frac{\omega _p^{2}}{{{\omega ^2}}},
\end{equation}
\begin{equation}\label{eq12}
{\epsilon_{\rm{J}}}\left( \omega  \right) = \epsilon_{\rm{J}}^\infty  + \frac{{f\omega _0^2}}{{\omega _0^2 - {\omega ^2}}}.
\end{equation}
Introducing new variable $x = {\left( {{\omega  \mathord{\left/
 {\vphantom {\omega  {{\omega _p}}}} \right.
 \kern-\nulldelimiterspace} {{\omega _p}}}} \right)^2}$ and the designation $\varkappa = {\left( {\omega  \mathord{\left/
 {\vphantom {\omega  {{\omega _p}}}} \right.
 \kern-\nulldelimiterspace} {\omega _{p}}} \right)^2}$, from the equation (\ref{eq10}) we obtain the following expression
\begin{eqnarray}\label{eq13}
&&\left( \epsilon_{\rm{J}}^\infty  + 2{\epsilon_{\rm{m}}} - \frac{{f\varkappa}}{{x - \varkappa}} \right)\left( \epsilon^{\infty } + 2\epsilon_{\rm{J}} - \frac{1}{x} - \frac{{2f\varkappa}}{{x - \varkappa}} \right) \nonumber \\
&&+ \beta _{\rm{c}}\left( \epsilon_{\rm{J}}^\infty  - {\epsilon_{\rm{m}}} - \frac{f\varkappa}{x - \varkappa} \right)\left(\epsilon^{\infty } - \epsilon_{\rm{J}}^\infty  - \frac{1}{x} + \frac{{f\varkappa}}{x - \varkappa} \right) = 0.
\end{eqnarray}
The expression (\ref{eq13}) can be presented in the form of the cubic equation
\begin{equation}\label{eq14}
{a_3}{x^3} + {a_2}{x^2} + {a_1}x + {a_0} = 0,
\end{equation}
where
\begin{eqnarray}\label{eq15}
  {a_0} &=&  - {\varkappa^2}\left[ 2\left( {1 - {\beta _{\rm{c}}}} \right){\epsilon_{\rm{m}}} + \left( 1 + 2\beta _{\rm{c}} \right)\left( f + \epsilon_{\rm{J}}^\infty  \right) \right], \\
  {a_1} &=& 2\left( 1 - {\beta _{\rm{c}}} \right)\left\{ {\left[ {\epsilon_{\rm{J}}^{\infty \,2} + {\epsilon_{\rm{m}}}{\epsilon^\infty } + f\left( {f + 2\epsilon_{\rm{J}}^\infty } \right)} \right]{\varkappa^2} + 2{\epsilon_{\rm{m}}}\varkappa} \right\} +  2\left( {2 + {\beta _{\rm{c}}}} \right){\epsilon_{\rm{m}}}\left( {f + \epsilon_{\rm{J}}^\infty } \right){\varkappa^2} \nonumber\\
  & + & \left( 1 + 2\beta _{\rm{c}} \right)  \left[ {\epsilon^{\infty }\left( {f + \epsilon_{\rm{J}}^\infty } \right){\varkappa^2} + \left( {f + 2\epsilon_{\rm{J}}^\infty } \right)\varkappa} \right], \\
  {a_2} &=&  - 2\left( {1 - {\beta _{\rm{c}}}} \right)\left[ {\epsilon_{\rm{m}} + 2\left( {\epsilon_{\rm{J}}^{\infty \,2} + f\epsilon_{\rm{J}}^\infty  + \epsilon_{\rm{m}}\epsilon^{\infty }} \right)\varkappa} \right] - \left( 1 + 2\beta _{\rm{c}} \right)\left[ {\epsilon_{\rm{J}}^\infty  + {\epsilon^\infty }\left( f + 2\epsilon_{\rm{J}}^\infty  \right)\varkappa} \right]\nonumber \\
  & - & 2\left( {2 + {\beta _{\rm{c}}}} \right){\epsilon_{\rm{m}}}  \left( {f + 2\epsilon_{\rm{J}}^\infty } \right)\varkappa, \\
    {a_3} &=& 2\left( {1 - {\beta _{\rm{c}}}} \right)\left( {\epsilon_{\rm{J}}^{\infty \,2} + {\epsilon_{\rm{m}}}{^\infty }} \right) + \left[ {\left( {1 + 2{\beta _{\rm{c}}}} \right){\epsilon^\infty } + 2\left( {2 + {\beta _{\rm{c}}}} \right){\epsilon_{\rm{m}}}} \right]\epsilon_{\rm{J}}^\infty,
\end{eqnarray}
which has the solution
\begin{eqnarray}\label{eq16}
  x_1 &=&  - \frac{{{a_2}}}{{3{a_3}}} + 2\sqrt { - \frac{p}{3}} \cos \frac{\beta }{3}, \\
  x_2 &=&  - \frac{{{a_2}}}{{3{a_3}}} - 2\sqrt { - \frac{p}{3}} \cos \left( {\frac{\beta }{3} + \frac{\piup }{3}} \right), \\
  x_3 &=&  - \frac{{{a_2}}}{{3{a_3}}} - 2\sqrt { - \frac{p}{3}} \cos \left( {\frac{\beta }{3} - \frac{\piup }{3}} \right),
\end{eqnarray}
where
\begin{equation}\label{eq17}
p =  - \frac{1}{3}{\left( {\frac{{{a_2}}}{{{a_3}}}} \right)^2} + \frac{{{a_1}}}{{{a_3}}},
\qquad
q = \frac{2}{{27}}{\left( {\frac{{{a_2}}}{{{a_3}}}} \right)^3} - \frac{{{a_1}{a_2}}}{{3a_3^2}} + \frac{{{a_0}}}{{{a_3}}},
\qquad
\cos \frac{\beta }{3} =  - \frac{q}{{2\sqrt { - {{\left( {{p}/{3}} \right)}^3}} }},
\end{equation}
and the frequencies of hybrid plasmon-exciton modes, which correspond to the solutions of the equation~(\ref{eq13}), are determined by the expressions
\begin{equation}\label{eq19}
{\omega _i} = {\omega _p}\sqrt {{x_i}},
\,\,\,\,
\left( {i = 1,\,2,\,3} \right).
\end{equation}
Subsequently, the relations (\ref{eq1})--(\ref{eq9})  and (\ref{eq19}), taking into account (\ref{eq10})--(\ref{eq17}), will be used to obtain the numerical results.

\section{The results of calculations and discussion}

The calculations were performed for the composite nanoparticles of the different sizes with the cores of different metals and the shells of different cyanine dyes, located in water ($\epsilon_m=1.77$) and in teflon ($\epsilon_m=2.3$). The parameters of metals and dyes are given in tables~\ref{tbl-smp1} and~\ref{tbl-smp2}.

\begin{table}[htb]
\caption{Parameters of metals ($a_0$ --- Bohr radius)(see, for example,~\cite{B70} and references therein).}
\label{tbl-smp1}
%\vspace{2ex}
\begin{center}
\renewcommand{\arraystretch}{0}
\begin{tabular}{|c||c|c|c|c|c|}
\hline
\raisebox{-1.7ex}[0pt][0pt]{Value}
      & \multicolumn{5}{c|}{Metals} \strut\\
\cline{2-6}
      & Cu& Au&  Ag&  Pt&  Pd \strut\\
\hline
${r_s}/{a_0}$ &2.11&3.01&3.02&3.27&4.00\strut\\
\hline
${m^*}/{m_e}$ &1.49&0.99&0.96&0.54&0.37\strut\\
\hline
${\epsilon^\infty }$ &12.03&9.84&3.70&4,42&2.52\strut\\
\hline
${\gamma _{{\rm{bulk}}}},\,\,{10^{14}}\,\,{{\rm{s}}^{-1}}$
&0.37&0.35&0.25&1.05&1.39\strut\\
\hline
\end{tabular}
\renewcommand{\arraystretch}{1}
\end{center}
\end{table}

\begin{table}[htb]
\caption{Parameters of J-aggregates (see, for example,~\cite{B35++} and references therein).}
\label{tbl-smp2}
%\vspace{2ex}
\begin{center}
\renewcommand{\arraystretch}{0}
\begin{tabular}{|c||c|c|c|}
\hline
\raisebox{-1.7ex}[0pt][0pt]{Value}
      &\multicolumn{3}{c|}{J-aggregates}\strut\\
\cline{2-4}
      & TC& OC&  PIC \strut\\
\hline
$\epsilon^{\inf}_{\rm{J}}$ &1&1&2.9\strut\\
\hline
$\hbar \omega_0$, eV &2.68&3.04&2.13\strut\\
\hline
$f$ &0.90&0.01&0.10\strut\\
\hline
$\hbar \gamma_{\rm{J}}$, eV&0.066&0.039&0.033\strut\\
\hline
\end{tabular}
\renewcommand{\arraystretch}{1}
\end{center}
\end{table}

\begin{figure}[htb]
\centerline{\includegraphics[scale=0.5]{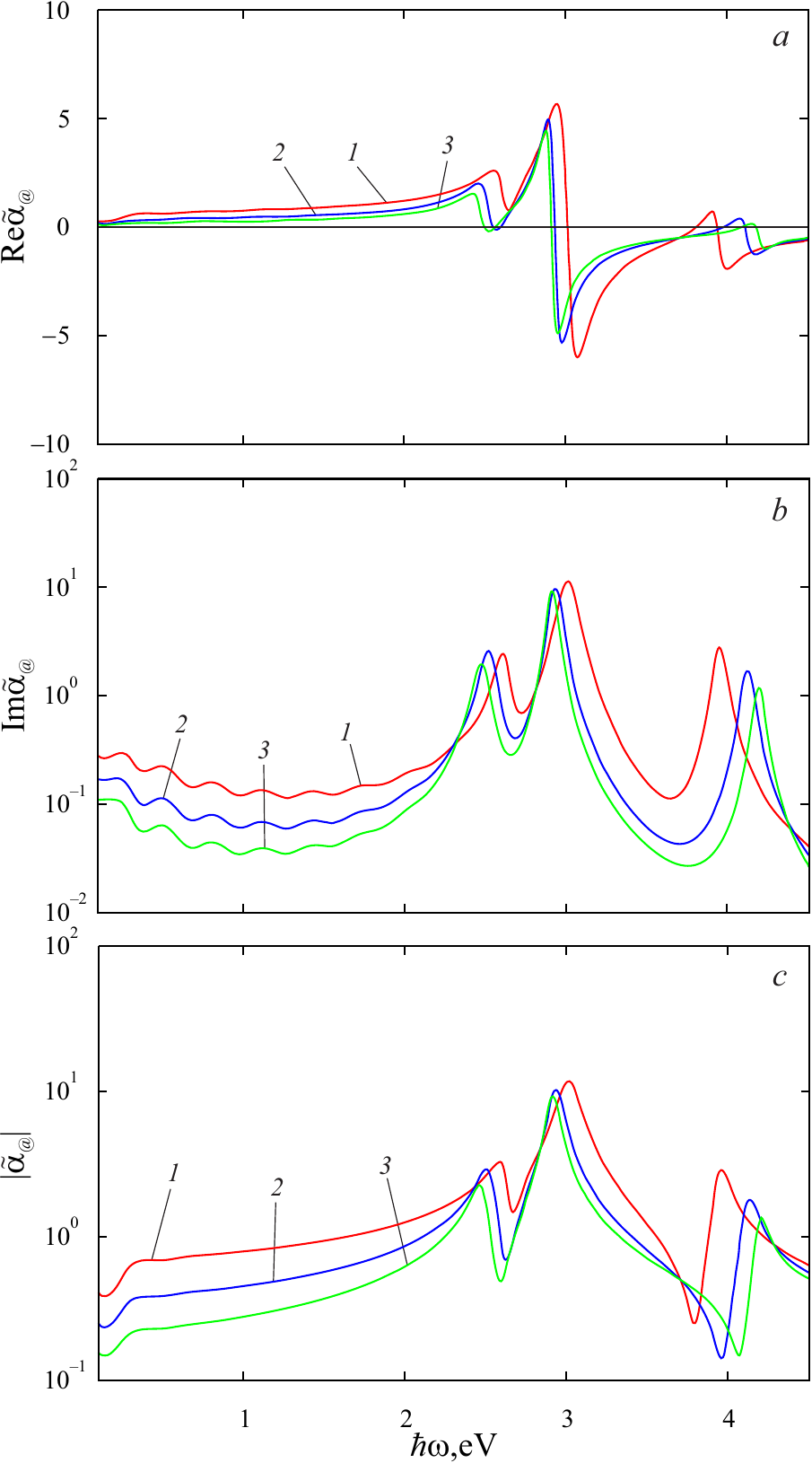}}
\caption{(Colour online) The frequency dependencies for the real (a) and imaginary (b) parts and for the module (c) of the dimensionless polarizability for the nanoparticles ${\rm{Ag}}@{\rm{TC}}$ with the fixed size of the core ($R_c=30{\rm{nm}}$) and different thickness of the shell: 1 -- $t = 1$ nm; 2 -- $t = 3$ nm; 3 -- $t = 5$ nm.}
\label{fig7}
\end{figure}

Figure~\ref{fig7} shows the results of the calculations of frequency dependencies for the real and imaginary parts and for the dimensionless polarizability of the nanoparticles ${\rm{Ag}}@{\rm{TC}}$ with the different thickness of the shell. An analysis of the obtained dependencies shows the presence of three maxima and three minimums $\Re \tilde \alpha _@$, $\Im {\tilde \alpha _@}$ and $\left| {{{\tilde \alpha }_@}} \right|$. At the same time, the second and the third maxima of $\Im {\tilde \alpha _@}$ are situated in the optical part of the spectrum, while the first maximum is situated in the ultraviolet region. Moreover, if $\Re {\tilde \alpha _@}({\hbar \omega })$ is the sign-variable function of frequency (figure~\ref{fig7}a) then $\Im {\tilde \alpha _@}({\hbar \omega }) > 0$ in the frequency range which is under the study (figure~\ref{fig7}b), and the values of the real and imaginary parts of polarizability are of the same order, hence the dependence $\left| {{{\tilde \alpha }_@}({\hbar \omega })} \right|$ possesses the features of both $\Re {\tilde \alpha _@}({\hbar \omega })$ (sharp minima) and $\Im {\tilde \alpha _@}({\hbar \omega })$ (close frequencies of maxima) (figure~\ref{fig7}c). In turn, the presence of the small-scale oscillations $\Im {\tilde \alpha _@}({\hbar \omega })$ in the infrared frequency range can be explained by the oscillating nature of the surface relaxation rate in this frequency range [see formulae~(\ref{eq7}) and~(\ref{eq8})]. Furthermore, there should be pointed out the presence of ``red'' shift of the second and the third maxima $\Im {\tilde \alpha _@}$ and ``blue'' shift of the first maximum under the increase in the thickness of the shell. Such “repulsion” of the surface plasmonic resonances for the spherical bimetallic nanoparticles was predicted in work~\cite{B70}.

\begin{figure}[htb]
\centerline{\includegraphics[scale=0.5]{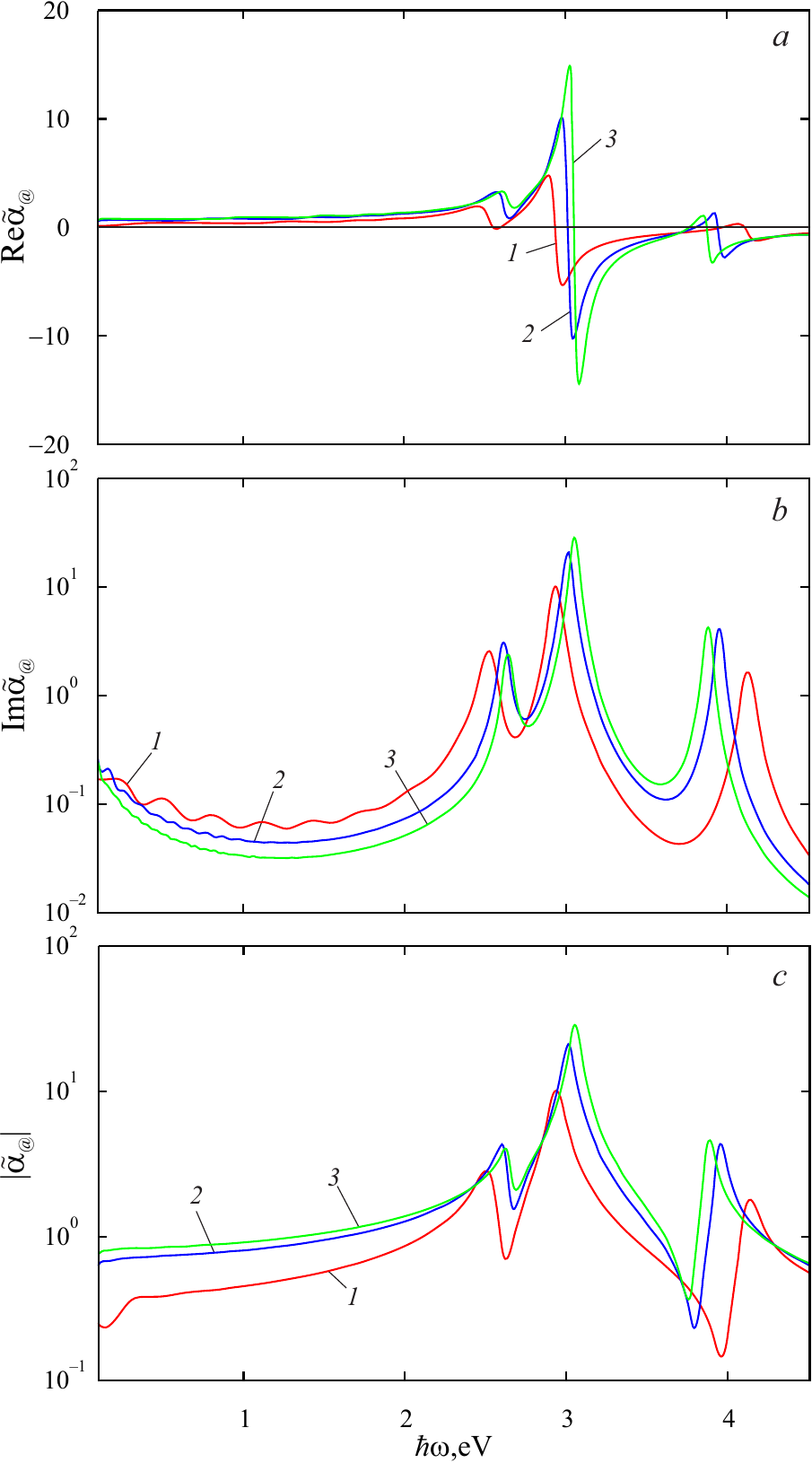}}
\caption{(Colour online) The frequency dependencies for the real (a) and imaginary (b) parts, and for the module (c) of the dimensionless polarizability for the nanoparticles ${\rm{Ag}}@{\rm{TC}}$ with the fixed thickness of the shell ($t = 3$~nm) under the different radius of the core: 1 --- $R_c = 10$ nm; 2 --- $R_c = 20$ nm; 3 --- $R_c = 30$~nm.}
\label{fig8}
\end{figure}

Figure~\ref{fig8} shows the graphs of the frequency dependencies $\Re {\tilde \alpha _@}$, $\Im {\tilde \alpha _@}$ and $\left| {{{\tilde \alpha }_@}} \right|$ for the nanoparticles ${\rm{Ag}}@{\rm{TC}}$ with the different size of the core. It should be pointed out that an increase of the radius of the core results in red shift of the first maxima and minimums $\Re {\tilde \alpha _@}$, $\Im {\tilde \alpha _@}$, $\left| {{{\tilde \alpha }_@}} \right|$ and blue shift of the second and the third maxima and minimums. The analysis of the nature of the shifts of the maxima of the frequency dependencies for the imaginary part of polarizability (figures~\ref{fig7}b and~\ref{fig8}b) indicates the presence of blue shift for the maxima from the optical frequency range and red shift of the maxima from ultraviolet region under an increase of the content of metal in the composite particle (increase of the radius of the metallic core or decrease of the thickness of the organic shell). In the limiting case $t \to 0$, all three maxima merge into one $\omega _{{\rm{res}}}^{\left( i \right)} \to {\omega _{sp}} = {{\omega_{p}} \mathord{/
 {\vphantom {{{\omega _p}} {\sqrt {\epsilon^{\infty } + 2\epsilon_{\rm{m}}} }}} 
 \kern-\nulldelimiterspace} {\sqrt {{\epsilon^\infty } + 2{\epsilon_{\rm{m}}}} }}$, which corresponds to SPR for the spherical metallic nanoparticle.

\begin{figure}[htb]
	\centerline{\includegraphics[scale=0.5]{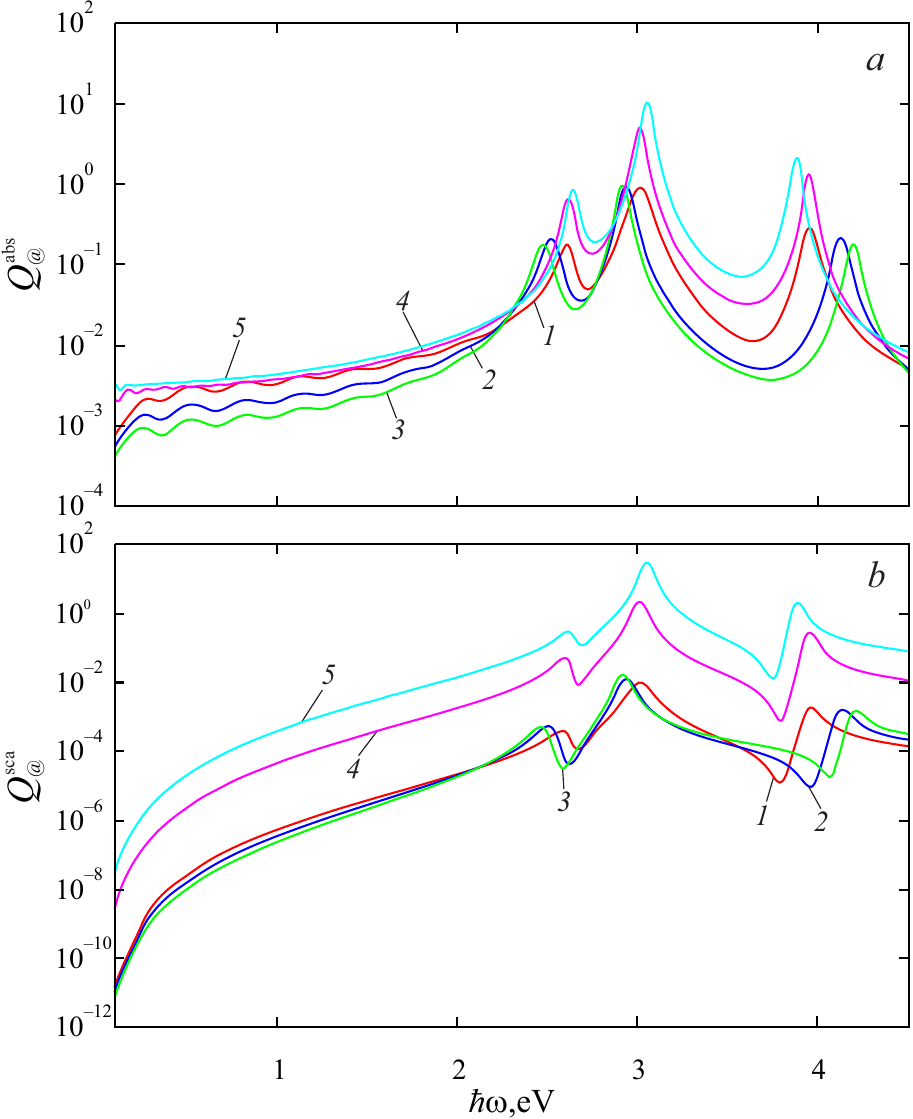}}
	\caption{(Colour online) The frequency dependencies of the absorption efficiencies (a) and scattering (b) for the nanoparticles ${\rm{Ag}}@{\rm{TC}}$ in teflon: \emph{1} --- $R_c = 30$ nm, $t = 1$ nm;
		\emph{2} --- $R_c = 30$ nm, $t = 3$ nm; \emph{3} --- $R_c = 30$ nm, $t = 5$ nm; \emph{4} --- $R_c = 10$ nm, $t = 3$ nm; \emph{5} --- $R_c = 50$ nm, $t = 5$ nm.}
	\label{fig9}
\end{figure}

Figure~\ref{fig9} shows the curves of the frequency dependencies for the absorption and scattering efficiencies for the nanoparticles ${\rm{Ag}}@{\rm{TC}}$. Since $Q_@^{\rm{abs}}\sim{\rm Im} {\tilde \alpha _@}$ and $Q_@^{{\rm{sca}}}\sim{\left| {{{\tilde \alpha }_@}} \right|^2}$, then the corresponding curves are qualitatively similar (the same number of extremes, the nature of their shifts under the variation of geometrical sizes and the presence of small-scale oscillations $\Im {\tilde \alpha _@}$ and $Q_@^{{\rm{abs}}}$ in the infrared part of the spectrum).

Now let us perform a comparison of the results obtained in the work for the spherical particles ${\rm
Me}@{\rm TC}$ with the available experimental results. It should be pointed out that the experimental studies of the optical properties of individual nanoparticles ${\rm Me}@{\rm TC}$ are a rather difficult task. Hence, due to this fact, they have not been carried out. That is why let us perform a qualitative comparison with the case of nanoparticles ${
\rm Ag}@{\rm TC}$ and ${\rm Au}@{\rm TC}$ in water solution~\cite{B35++}. Such a comparison is qualitative because the experiment was carried out in the wavelength range $\lambda =300\div 800$~nm and the photoabsorption was measured in relative units, while our work represents the frequency dependencies for the absorption efficiencies. Thus, the measurements in the indicated wavelength range did not allow us to detect maximum in the near ultraviolet range, which corresponds to $\lambda _{\max }^{\left(1\right)} \approx 250$~nm (or $\hbar \omega _{{\rm res}}^{\left(1\right)}
=5.11$~eV for ${\rm Au}@{\rm TC}$ and $\hbar \omega _{{\rm res}}^{\left(1\right)}
=5.03$~eV for ${\rm Ag}@{\rm TC}$), as estimated according to formulae
(\ref{eq13})--(\ref{eq19}) under $\beta _{\rm c} \approx 0.6$ [$\beta _{\rm c} =\left({t\mathord{\left/ {\vphantom {t R}}
\right. \kern-\nulldelimiterspace} R} \right)^{3} $, $t=1$~nm, $R_{{\rm c}} =5$~nm
are the sizes of colloidal particles in the experiment]. As for the second maximum of the experimental curve ($\lambda _{\max }^{\left(2\right)} \approx 400$~nm for ${\rm Ag}@{\rm TC}$ and $\lambda
_{\max }^{\left(2\right)} \approx 420$~nm for ${\rm Au}@{\rm TC}$), then $\hbar \omega
_{{\rm res}}^{\left(2\right)} =2.97$~eV for ${\rm Ag}@{\rm TC}$ and $\hbar \omega
_{{\rm res}}^{\left(2\right)} =2.75$~eV for ${\rm Au}@{\rm TC}$ correspond to $\lambda
_{\max }^{\left(2\right)} \approx 450$~nm and $\lambda _{\max }^{\left(2\right)}
\approx 420$~nm, i.e., we have almost perfect match of the resonant wavelength for the particles ${\rm Au}@{\rm TC}$. For the third maximum ($\lambda _{\max }^{
\left(3\right)} \approx 480$~nm for ${\rm Ag}@{\rm TC}$ and $\lambda _{\max }^{\left(3
\right)} \approx 550$~nm for ${\rm Au}@{\rm TC}$), and $\hbar \omega _{{\rm res}}^{
\left(3\right)} =2.11$~eV for ${\rm Ag}@{\rm TC}$ and $\hbar \omega _{{\rm res}}^{
\left(3\right)} =1.62$~eV for ${\rm Au}@{\rm TC}$, which correspond to $\lambda _{
\max }^{\left(3\right)} \approx 765$~nm and $\lambda _{\max }^{\left(3\right)} \approx
590$~nm, i.e., the theoretical and experimental values $\lambda _{\max }^{\left(3\right)} $ are close for ${\rm Au}@{\rm TC}$ while the values
$\lambda
_{\max }^{\left(3\right)} $ are significantly different for
${\rm Ag}@{\rm TC}$ the values differ significantly. In our opinion, this can be associated with the fact that the theoretical values of resonant frequencies $\hbar \omega
_{{\rm res}}^{\left(i\right)} $ were estimated in non-dissipative approximation, while under an increase of the wavelength (decrease of the frequency) of the incident light for the particles of the indicated size, the relaxation processes, associated with the electron scattering both by the surface of the particle and with the additional chemical damping at the interface ``metal -- J-aggregate'' interface (see, for example, \cite{B71,B72}), may be essential. Nevertheless, the authors are aware that this issue requires a more detailed research and will be the subject of subsequent works of the authors.

\begin{figure}[htb]
\centerline{\includegraphics[scale=0.55]{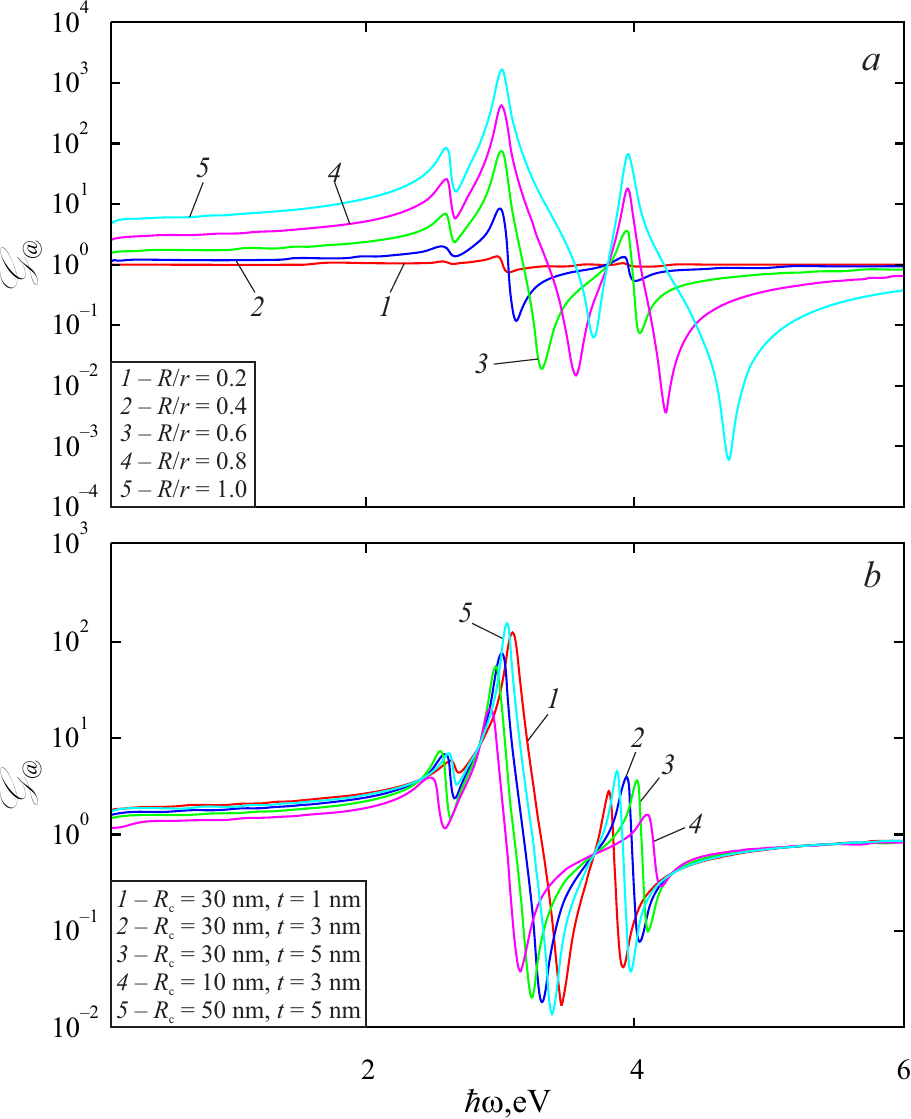}}
\caption{(Colour online) The frequency dependencies for the amplification factor for the nanoparticles Ag@TC in teflon under a fixed size ($R_c = 30$ nm, $t = 3$ nm) at a different distance from its surface (a) and under a fixed distance ($R \diagup r$ = 0.6) from the surface under different sizes (b).}
\label{fig10}
\end{figure}

Figure~\ref{fig10} shows the frequency dependencies for the amplification of the electric field. It is clear that an increase of the distance from the surface of the nanoparticle results in a sharp decrease of the amplification magnitude. The amplification is maximum near the surface of the nanoparticle. In turn, the variations of the values and the shifts of the amplification maxima under the variation of the core sizes and the thickness of the shell are similar to the behavior of the frequency dependencies for the polarizability module in full correspondence with the expression (\ref{eq9}).

\begin{figure}[htb]
	\centerline{\includegraphics[scale=0.55]{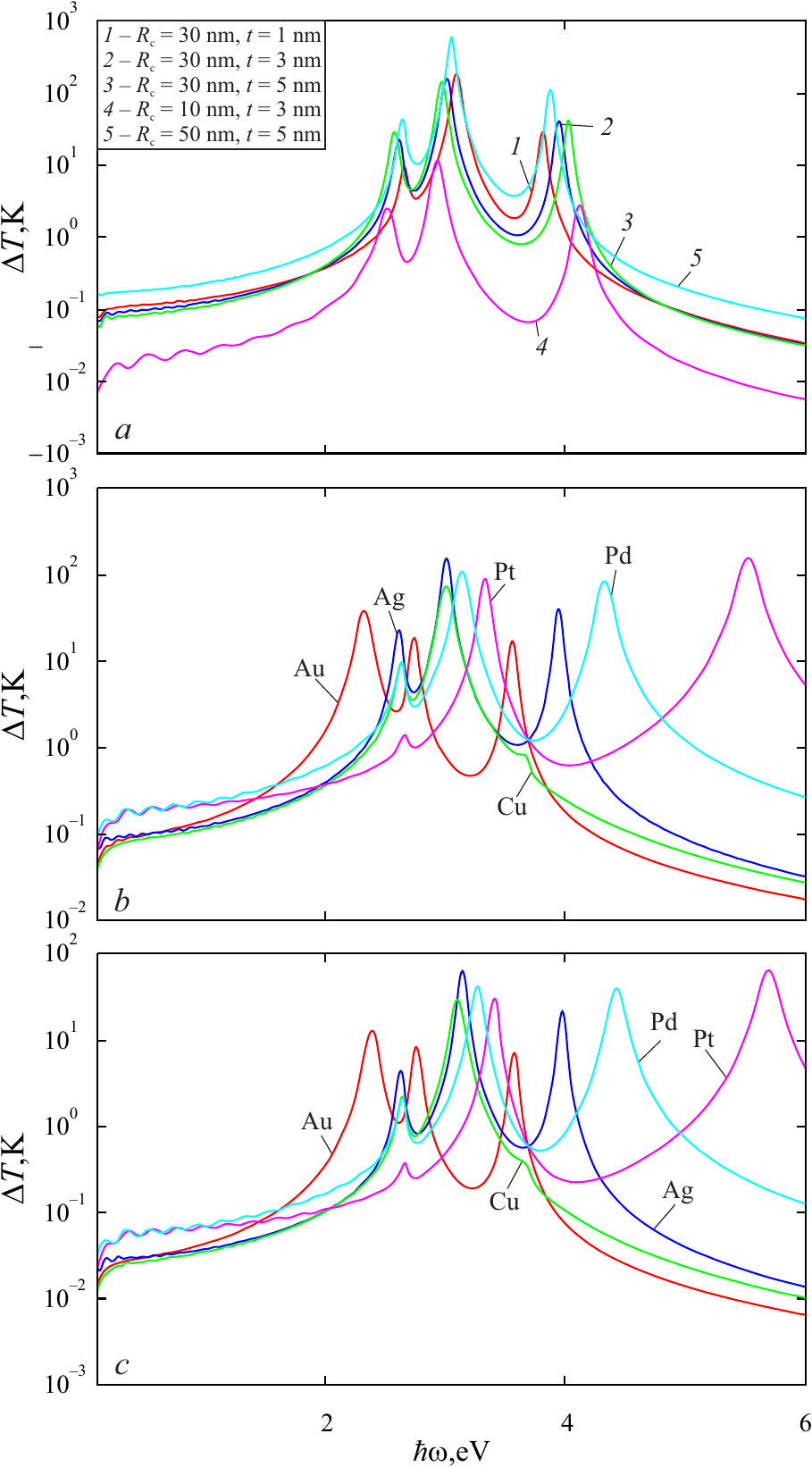}}
	\caption{(Colour online) The frequency dependencies for the heating of the nanoparticles ${\rm{Ag}}@{\rm{TC}}$ of different sizes in teflon (a); nanoparticles ${\rm{Me}}@{\rm{TC}}$ of fixed size ($R_c = 30$ nm, $t = 3$ nm) in teflon (b) and in water (c).}
	\label{fig11}
\end{figure}

Figure~\ref{fig11} shows the frequency dependencies for the temperature heating for the nanoparticles ${\rm{Ag}}@{\rm{TC}}$ of different sizes and nanoparticles of different content in different dielectric media. The curves $\Delta T({\hbar \omega })$ for the nanoparticles of different sizes (figure~\ref{fig11}a) are qualitatively similar to the analogous curves $Q_@^{{\rm{abs}}}({\hbar \omega })$ both by the nature of the shifts and by the relative values of maxima, as well as by the presence of small-scale oscillations in the infrared part of the spectrum, because $\Delta T\sim Q_@^{\rm{abs}}$. It should be pointed out that the location and the value of the maxima of the heating depend essentially on the core material and the environment, which is due to the essential differences between plasma frequencies ${\omega _p}$ for different metals and permittivity of the dielectric environment (figure~\ref{fig11}b and~c). Thus, the maximum values of heating for the nanoparticles in teflon are approximately two times as much as the values of heating in water. Moreover, small ``blue'' shift of $\Delta {T_{\max }}$ takes place for the nanoparticles in water with respect to $\Delta {T_{\max }}$ for the particles in teflon.

\begin{figure}[htb]
	\centerline{\includegraphics[scale=0.53]{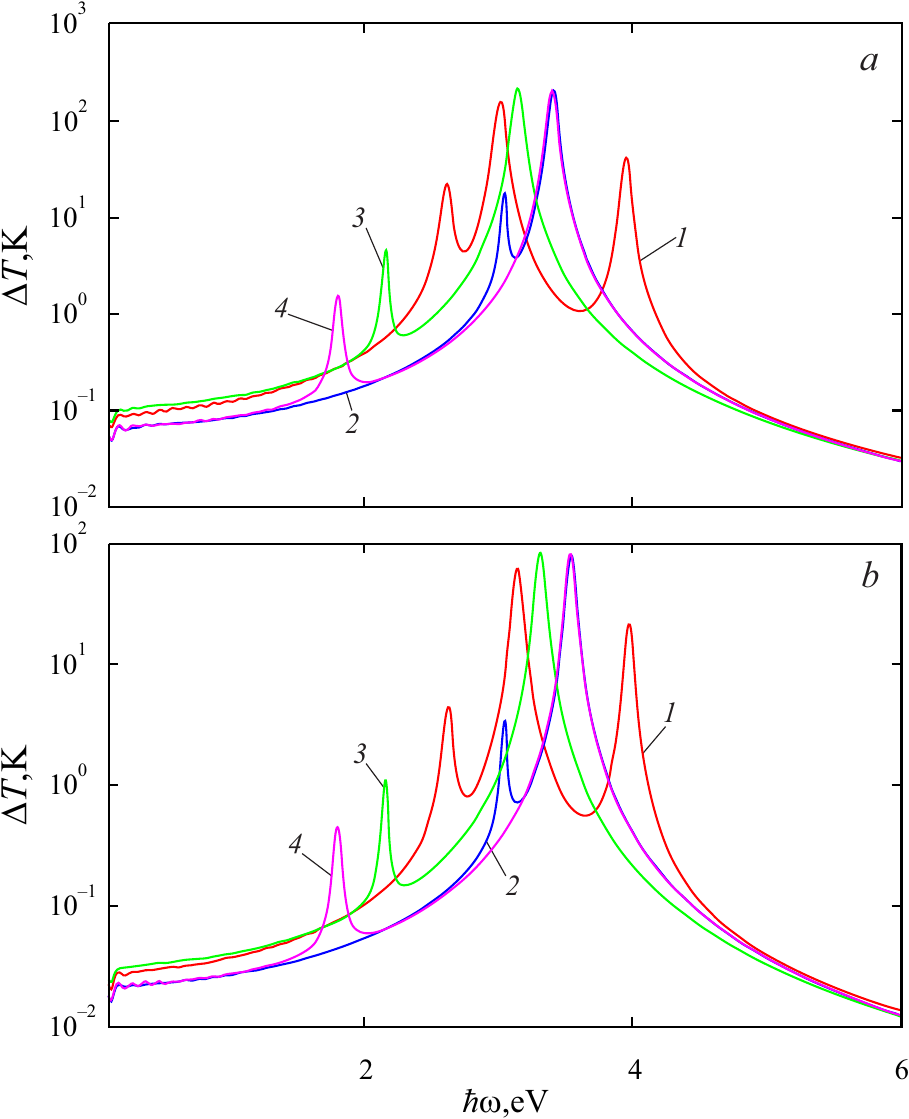}}
	\caption{(Colour online) The frequency dependencies for the heating for the nanoparticles ${\rm{Ag}}@{\rm{J}}$ of a fixed size ($R_c = 30$ nm, $t = 3$ nm) in teflon (a) in water (b) with different material of the shell: 1 --- TC; 2 --- OC; 3 --- PIC; 4 --- NK2567.}
	\label{fig12}
\end{figure}

Figure~\ref{fig12} shows the frequency dependencies for the heating of the nanoparticles with the core Ag and with the shells of different cyanine dye in different media. The nature of the curves for the cases of the particles ${\rm{Ag}}@{\rm{J}}$ in teflon and water (the presence of three maxima as well as their location in different frequency ranges) is similar. The difference between the indicated curves is only in great values $\Delta {T_{\max }}$ for the nanoparticles in teflon.

Let us now consider the question of the expediency of the application of the particles ${\rm{Me}}@{\rm{J}}$ in nanomedicine, in particular, for photothermal therapy of malignant neoplasms. For this purpose, we use the concepts developed in work~\cite{B41}. It was established in this work that the thermal conductivity process in tumor tissues can be considered to be quasi-stationary, and the temperature distribution in tumor tissues is determined by the relation (it is assumed that the tumor has a spherical form)
\begin{equation}
  T\left( r \right) = {T_0} + \frac{I_{0}}{k_{a}\lambda }\left[ \left( 1 +k_{a}r \right){{\mathop{\rm e}\nolimits} ^{ - {k_a}r}} - \left( 1 +k_{a}R_{t} \right){{\mathop{\rm e}\nolimits} ^{ -k_{a}R_{t}}} \right],
\end{equation}
where $T_0$ is the temperature of environment; $k_{a} = C_@^{\rm{abs}}{N_T} +k_{t}$ is the absorption coefficient (${N_T} = {10^{12}}\,\,{\rm{m}}{{\rm{l}}^{ - 1}}$, ${k_t} = 8\,\,{{\rm{m}}^{ - 1}}$~\cite{B41}); $R_t$ is the radius of tumor.

The calculations for metal-graphene particle show that it is necessary either to move the particles inside the tumor or to use conglomerate of nanoparticles in order to achieve the heating which is sufficient for tumor necrosis~\cite{B41}. At the same time, the calculations for the bimetallic nanoparticles indicate a strong overheating in their neighborhood, which makes their use impossible under the therapy of the malignant tumors~\cite{B42,B43}. Hence, the performance of similar calculations for ${\rm{Me}}@{\rm{J}}$ particles is of great interest.

\begin{figure}[htb]
\centerline{\includegraphics[scale=0.55]{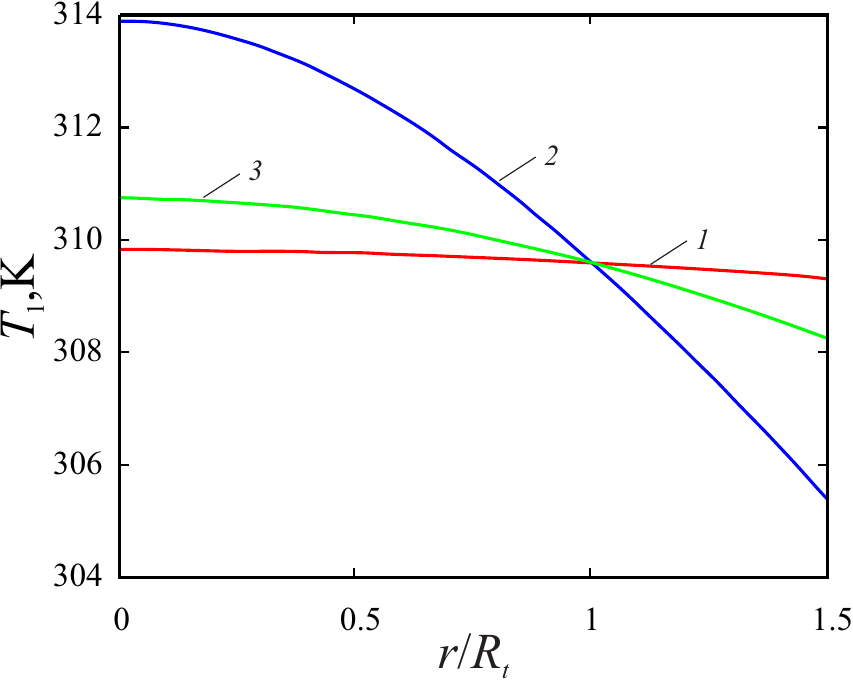}}
\caption{(Colour online) The temperature distributions, calculated under the frequencies of hybrid plasmon-exciton resonances: 1 --- $\omega=\omega_{\rm{res}}^{\left(1\right)}$; 2 --- $\omega=\omega_{\rm{res}}^{\left(2\right)}$; 3 --- $\omega=\omega_{\rm{res}}^{\left(3\right)}$. }
\label{fig13}
\end{figure}

Figure~\ref{fig13} shows the radial distribution of the temperature in the malignant neoplasm tissues at the frequencies of the hybrid plasmon-exciton resonances. Thermophysical parameters of the healthy tissue and tumor tissue are given in table~\ref{tbl-smp3}. It should be pointed out that a decrease of temperature with distance from the center of the tumor takes place for all resonance frequencies. However, the temperatures, which are sufficient for the necrosis of the malignant tissues (provided that healthy tissues are not damaged, i.e., $t < 42$\textcelsius), are reached only under $\omega=\omega_{\rm{res}}^{\left(2\right)}$. This is confirmed by the fact that the maximum heating of the composite nanoparticles takes place under $\omega=\omega_{\rm{res}}^{\left(2\right)}$. Moreover, the temperatures, sufficient for a thermal destruction of the tumor, are reached practically through the whole volume of the tumor (up to $r = R_t$) under $\omega=\omega_{\rm{res}}^{\left(2\right)}$. The comparison of the calculation results for bimetallic and metal-graphene nanoparticles, as well as nanoparticles ``metallic core – J-aggregate shell'' indicates that the practical application of the latter is more expedient.

\begin{table}[htb]
\caption{The parameters of body tissue~\cite{B41}.}
\label{tbl-smp3}
%\vspace{2ex}
\begin{center}
\renewcommand{\arraystretch}{1.22}
\begin{tabular}{|c||c|c|}
\hline
\raisebox{-1.7ex}[0pt][0pt]{Parameter}
      &\multicolumn{2}{c|}{Value}\strut\\
\cline{2-3}
      & The medium 1 (tumor)& The medium 2 (healthy tissue)\strut\\
\hline
$\rho ,\,\,\frac{{{\rm{kg}}}}{{{{\rm{m}}^{\rm{3}}}}}$ &1080&1050\strut\\
\hline
$\lambda ,\,\,\frac{{\rm{W}}}{{{\rm{m}} \cdot {\rm{K}}}}$ &0.54&0.58\strut\\
\hline
${w_b},\,\,\frac{{{\rm{kg}}}}{{{{\rm{m}}^{\rm{3}}} \cdot {\rm{s}}}}$ &0.689&0.540\strut\\
\hline
${Q_{m0}},\,\,\frac{{\rm{W}}}{{{{\rm{m}}^{\rm{3}}}}}$&0.52&0.52\strut\\
\hline
$c,\,\,\frac{{\rm{J}}}{{{\rm{kg}} \cdot {\rm{K}}}}$&3700&3700\strut\\
\hline
\end{tabular}
\renewcommand{\arraystretch}{1}
\end{center}
\end{table}

\section{Conclusions}

The expressions for polarizability of the spherical nanoparticles with the structure ``metallic core~-- J-aggrerate shell'', the absorption and scattering efficiencies, heating of the nanoparticles and the amplification of the field in their neighborhood are obtained taking into account the frequency dependence for the dielectric permittivity of cyanine dyes.

It is established that the frequency dependence for the real part of the polarizability has three maxima, which correspond to three hybrid plasmon-exciton resonances. Moreover, two of them are located in the visible (optical) part of the spectrum and one of them --- in the ultraviolet part. Comparison of the results of theoretical calculations of resonances with experimental data showed good agreement for the case of particles ${\rm Au}@{\rm TC}$.

It is shown that an increase of content of the metallic fraction in the composite nanoparticle (an increase of the radius of metallic core or a decrease of the thickness of J-aggregate shell) results in red shift of the maximum from ultraviolet frequency range and in blue shift of the maxima from the visible part of the spectrum. The presence of small-scale oscillations of the imaginary part and module of polarizability in the infrared part of the spectrum is explained by the oscillations of the surface relaxation rate at these frequencies.

We demonstrated a qualitative similarity of the curves of the frequency dependencies for the absorption and scattering efficiencies, amplification of the fields in the neighborhood of the particle and its temperature heating to the curves for the imaginary part and module of polarizability.

It is established that the material of metallic core and shell as well as the properties of the environment have a significant effect on the location and value of the maxima of the heating of the nanoparticle.

The possible conditions of the use of the considered composite nanoparticles in photothermal therapy of the malignant tumors are determined. It is shown that the use of nanoparticles, which are under consideration, is more advisable than the use of metal-graphene or bimetallic particles, since, in practice, it is easier to achieve the necessary heating through the whole volume of the malignant tumor.

\ukrainianpart

\title{Оптичні і теплові ефекти в околі сферичної шаруватої наночастинки зі структурою ``металеве ядро -- оболонка J-агрегату''}
\author{А.~В.~Коротун\refaddr{label1,label2}, Н.~А.~Смирнова\refaddr{label1}, В.~І.~Рева\refaddr{label1}, І.~М.~Тітов\refaddr{label3}, Г.~М.~Шило\refaddr{label4}}
\addresses{
\addr{label1} Національний університет «Запорізька політехніка», вул. Жуковського, 64,
Запоріжжя, 69063, Україна
\addr{label2} Інститут металофізики ім. Г.~В.~Курдюмова НАН України, бульв. Академіка Вернадського, 36,
  Київ, 03142, Україна
\addr{label3} UAD Systems, вул. Олександрівська, 84, Запоріжжя, 69002, Україна
\addr{label4} Запорізький національний університет, вул. Жуковського, 66, Запоріжжя, 69600, Україна
}
%
%% якщо автор є один або автори є з однієї установи:
%
%  \author{1й Автор, 2й Автор, \ldots}
%  \address{Інститут\ldots}
%
%%

\makeukrtitle

\begin{abstract}
\tolerance=3000%
У роботі отримано співвідношення для поляризовності металевих наночастинок, вкритих оболонкою ціанінових барвників.
Досліджено частотні залежності ефективностей поглинання та розсіювання світла, розігріву композитної наночастинки та
підсилення електричного поля в її околі. Встановлено, що всі залежності мають три максимуми, які відповідають частотам
гібридного плазмон-екситонного резонансу. Показано, що збільшення вмісту металу в наночастинці призводить до синього
зміщення максимумів із видимої області спектра і червоного зміщення максимуму з ультрафіолетового діапазону частот.
Дос\-лід\-же\-но питання щодо застосування метал-органічних наночастинок у наномедиціні, зокрема, для фототермічної терапії
злоякісних новоутворень.
\keywords композитні наночастинки, J-агрегат, плазмон-екситонні резонанси, поляризовність, ефективності поглинання та розсіювання

\end{abstract}

\end{document}